\documentclass[11pt,a4paper]{article}
\usepackage[utf8]{inputenc}
\usepackage[T1]{fontenc}
\usepackage{amsmath,amssymb}
\usepackage{graphicx}
\usepackage{caption}
\usepackage{tabularx}
\usepackage{booktabs}
\usepackage{geometry}
\usepackage{natbib}
\usepackage{float}                
\usepackage{tabularx}             
\usepackage{array}                
\newcolumntype{C}{>{\centering\arraybackslash}X}  

\usepackage{hyperref}
\hypersetup{
  colorlinks=true,
  linkcolor=blue,
  citecolor=blue,
  urlcolor=blue
}

\title{Identifying High-Risk Areas for Traffic Collisions in Montgomery, Maryland\\Using KDE and Spatial Autocorrelation Analysis}
\author{Stanislav Liashkov\\
M.S.\ Candidate, Data Science\\
University of Colorado Boulder\\
\texttt{stanislav.liashkov@colorado.edu}}
\date{}  

\begin{document}
\maketitle

\begin{abstract}
{Despite a global decline in motor vehicle crash fatalities due to improved research and road safety policies, road traffic injuries remain a significant public health concern. The World Health Organization's 2023 report highlights that road traffic injuries are the leading cause of death among individuals aged 5-29, with over half of fatalities involving pedestrians, cyclists, and motorcyclists. This study addresses this critical issue by identifying high-risk areas in Montgomery County, Maryland, contributing to the global goal of halving road traffic deaths and injuries by 2030. Using Kernel Density Estimation (KDE) and spatial autocorrelation analysis, we estimate collision densities and identify hotspots for targeted interventions. Our findings reveal significant spatial clustering of traffic collisions, with distinct patterns in densely populated urban areas and rural regions, offering valuable insights for policymakers to enhance road safety.}
\end{abstract}


\section{Introduction}
Montgomery County stands as the most populous county in the U.S. state of Maryland, with a population of 1,062,061 as recorded in the 2020 census, reflecting a 9.3\% growth since 2010. The county seat is Rockville, while Germantown holds the distinction of being the most populous municipality within the county. Situated adjacent to Washington, D.C., Montgomery County forms a vital part of both the Washington metropolitan area and the Washington--Baltimore combined statistical area. The majority of its residents live in bustling communities such as Silver Spring, Bethesda, Germantown, and the incorporated cities of Rockville and Gaithersburg.

Motor vehicle crashes are a critical public health concern, particularly for young individuals, for whom they are the leading cause of death in the United States. Developing effective strategies to mitigate these fatalities requires a comprehensive understanding of the spatial and temporal patterns of traffic collisions. Traditional statistical approaches often fail to account for the complex spatial relationships inherent in collision data, underscoring the need for advanced spatial analysis methods to identify and prioritize the most hazardous hotspots for policymakers.

In spatial analysis, Moran's I is utilized to evaluate whether the distribution of collision counts deviates from spatial randomness, providing a global measure of spatial dependence. Local Moran's I complements this by identifying statistically significant clusters of spatial autocorrelation and classifying them based on their characteristics.

LISA clustering complements this by identifying localized hotspots and assessing the significance of spatial autocorrelation for specific locations. By calculating Local Moran's I for each census tract, LISA categorizes locations into cluster types, such as high-high (HH) clusters, where the target variable is relatively high and surrounded by similar areas, and low-low (LL) clusters, where the target variable is relatively low and similarly clustered. Importantly, only locations with significant spatial autocorrelation, as determined by pseudo p-values from Local Moran's I, are considered meaningful.

In addition, Kernel Density Estimation (KDE) is employed to visualize collision densities across Montgomery County. KDE generates heat maps that highlight areas with high concentrations of collisions, facilitating the identification of critical zones for effective interventions.

In this work, we apply these techniques to analyze traffic collisions in Montgomery County. Our objective is to identify and highlight unsafe areas prone to collisions that may require targeted interventions to enhance road safety and reduce fatalities.

\section{Related Work}
The problem of finding hotspots of road crashes has been researched for decades and several approaches have been established. Though there is no unique standard of approaching this kind of problem, spatial statistics methods have been gaining popularity among researchers in comparison to other methods.

Dezman et al.~\citep{ref-dezman} analyzed traffic collisions in Baltimore from 2009 to 2013, utilizing ARIMA modeling to identify temporal patterns and spatial autocorrelation techniques, such as Moran's I~\citep{ref-moran} and LISA~\citep{ref-lisa} clustering, to pinpoint crash hotspots. Additionally, they employed Kernel Density Estimation (KDE) to illustrate dense hotspots for various types of collisions. Their study revealed that most collisions occurred in high-density urban areas and at intersections with significant pedestrian activity, with distracted driving emerging as the dominant risk factor. These findings underscore the importance of addressing modifiable factors, such as road design and driver behavior, to enhance safety. Building upon their work, our geospatial analysis applies similar methodologies to Maryland's collision data from 2015 to 2024. By utilizing their approach in a different region and larger time frame, we aim to uncover insights into traffic collision patterns within Montgomery country.

Researches Ali Soltani et al.~\citep{ref-soltani} have applied methods from spatial statistics such as Getis Ord Gi*~\citep{ref-getis} statistics and Moran's I to detect hotspots of road crashes in Shiraz city and found that collisions produced significant clustering. The result of their study confirmed crashes hotspots generally appeared on arterial roadway which tend to have higher car speed/volume and more travel lanes. Md Saiful Alam et. al~\citep{ref-alam} utilized Getis Ord Gi*, the crash severity index and Moran's I to evaluate distribution of road traffic crashes in Ohio. As result of their research, they discovered significant hotspots in big cities of state: Cleveland, Cincinnati, Toledo, and Columbus and demonstrated the efficiency of spatial analysis applied to identifying dangerous areas prone to vehicle crashes.

\section{Data}

The dataset was compiled from open data on traffic collisions in Montgomery County, Maryland.~\citep{ref-montgomerycounty} These data were collected by the Automated Crash Reporting System (ACRS)~\citep{ref-acrs} of the Maryland State Police and reported by the Montgomery County Police, Gaithersburg Police, Rockville Police, and Maryland-National Capital Park Police. The dataset includes general information about each collision as well as detailed records of all incidents reported within the county. Various characteristics and contributing factors are documented, including location, vehicle type, type of harmful event, driver distraction, alcohol involvement, and more.The dataset can be found here~\citep{ref-dataset}.

\subsection{Scope of Data}

This dataset focuses on Montgomery County in Maryland (figure 1), USA, spanning the years 2015 to 2024. Over this period, around 106,000 road accidents were documented, including 2,547 severe incidents that caused fatalities or serious injuries as result of collision.

\begin{figure}[H]
\centering
\includegraphics[width=0.6\textwidth]{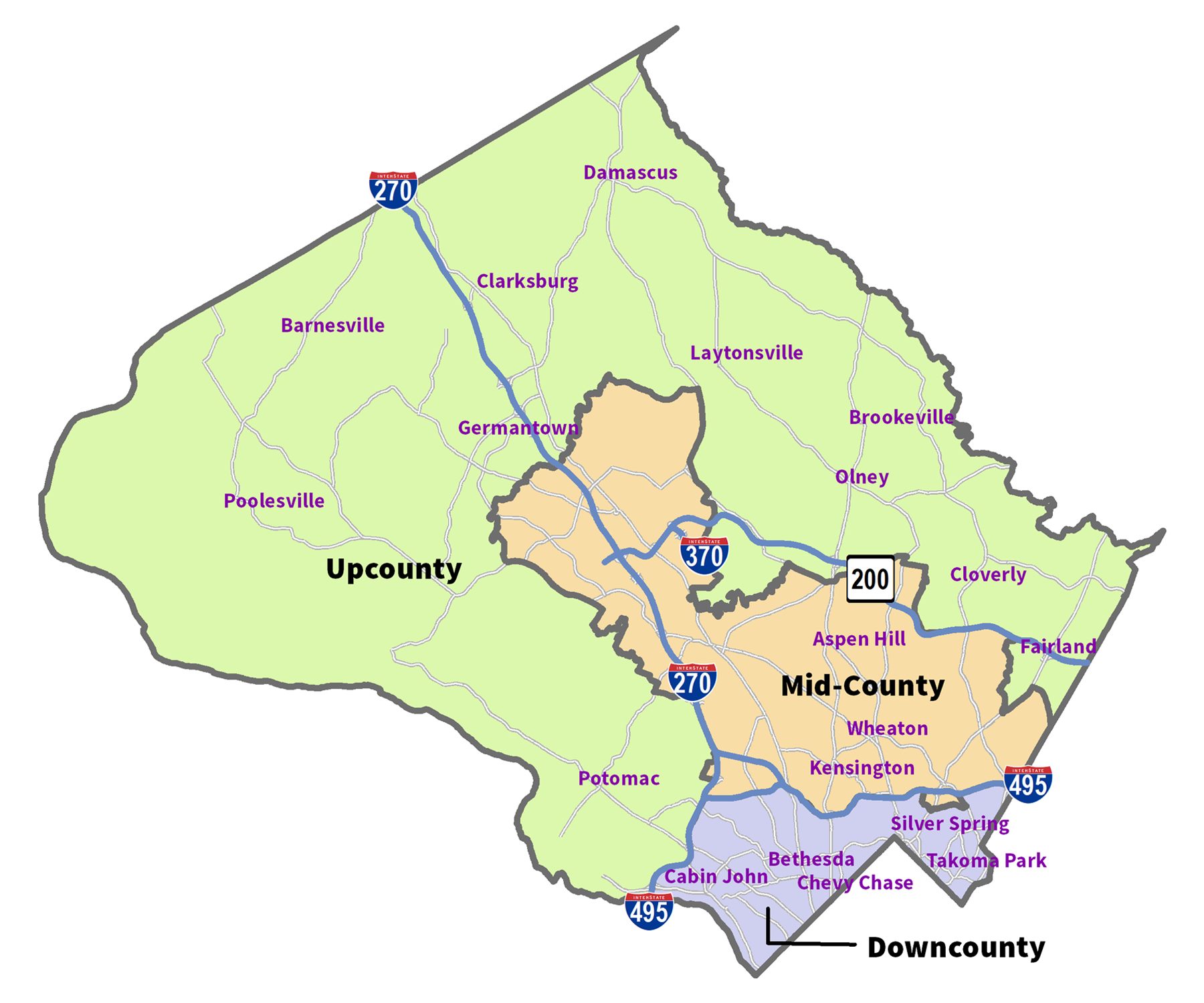}
\caption{Montgomery Country, Maryland - Planning map (link)}
\label{fig:montgomery_map}
\end{figure}

\subsection{Schema}

The dataset consists of three interconnected tables: incidents, drivers, and non-motorists. The incidents table provides general information about each collision, including time, location, road conditions etc. Incidents may involve multiple drivers, detailed in the drivers table, and occasionally non-motorists, such as pedestrians or cyclists, described in the non-motorists table

\begin{table}[H]
\caption{Metadata of tables in Dataset (after normalization)}
\label{tab:metadata}
\centering
\begin{tabularx}{\textwidth}{XCCC}
\toprule
\textbf{Table name} & \textbf{Number of rows} & \textbf{Number of columns} & \textbf{Entity} \\
\midrule
incidents & 106k & 17 & reported collision \\
drivers & 180k & 40 & driver involved \\
non-motorist & 6k & 14 & pedestrian/cyclist involved \\
\bottomrule
\end{tabularx}
\end{table}

\begin{figure}[H]
\centering
\includegraphics[width=0.8\textwidth]{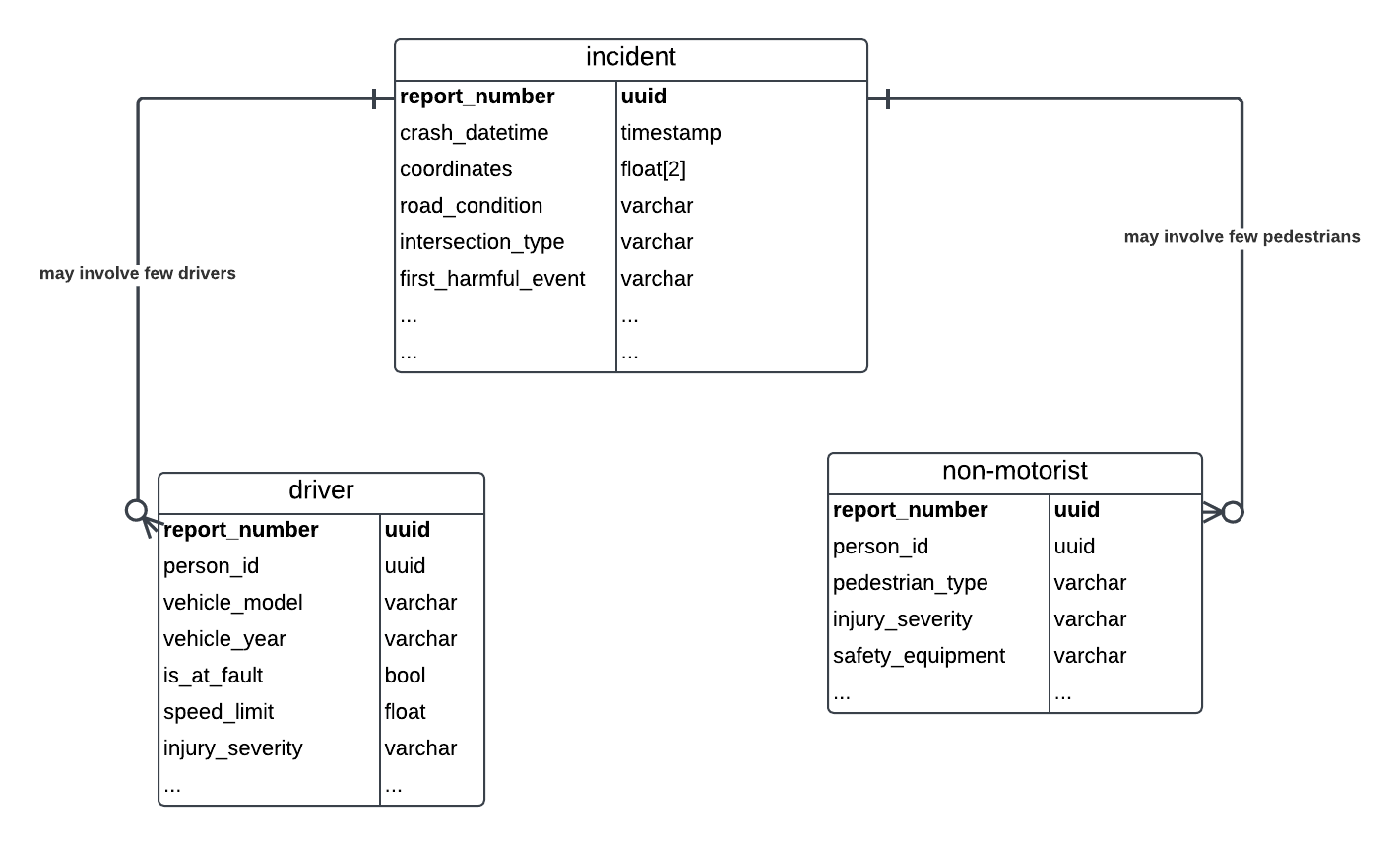}
\caption{Entities Relationship Diagram (ERD)}
\label{fig:erd}
\end{figure}

\subsection{Limitations of Data}

The collision data used in this study is subject to several limitations that should be carefully considered. As specified on the county's data provision website, collision reports are based on preliminary information provided to the Police Department by the reporting parties. This approach may introduce potential challenges, including:

\begin{itemize}
\item \textbf{Unverified Information:} Initial reports may include details that have not been corroborated through further investigation, potentially leading to inaccuracies.
\item \textbf{Incomplete Collision Details:} Certain elements of collisions may be missing or inaccurately recorded in the preliminary reports, affecting data completeness.
\item \textbf{Subsequent Updates:} Collision data may be revised or updated following official investigations, resulting in inconsistencies and potential discrepancies within the dataset.
\item \textbf{Reporting Errors:} Mechanical malfunctions or human errors during data entry and reporting processes may lead to inaccuracies in the collision records.
\end{itemize}

\section{Results}
\subsection{Time Series}

The time series plot (Figure 2) displays evident seasonality in monthly collision incidents along with an overarching trend. Notably, in the spring of 2020, there is a pronounced structural break and a significant reduction in traffic accidents attributed to the COVID-19 pandemic and associated lockdown measures. This deviation beginning in spring 2020 and extending into subsequent months makes the time series more complex for modeling with traditional approaches such as ARIMA or SARIMA, as it violates the assumptions of stationarity and homoscedasticity. Furthermore, the plot indicates that the overall trend in the total number of collisions is slowly recovering toward pre-pandemic levels. Nevertheless, even several years after COVID-19, the number of collisions remains below pre-pandemic figures.

\begin{figure}
\centering
\includegraphics[width=0.9\textwidth]{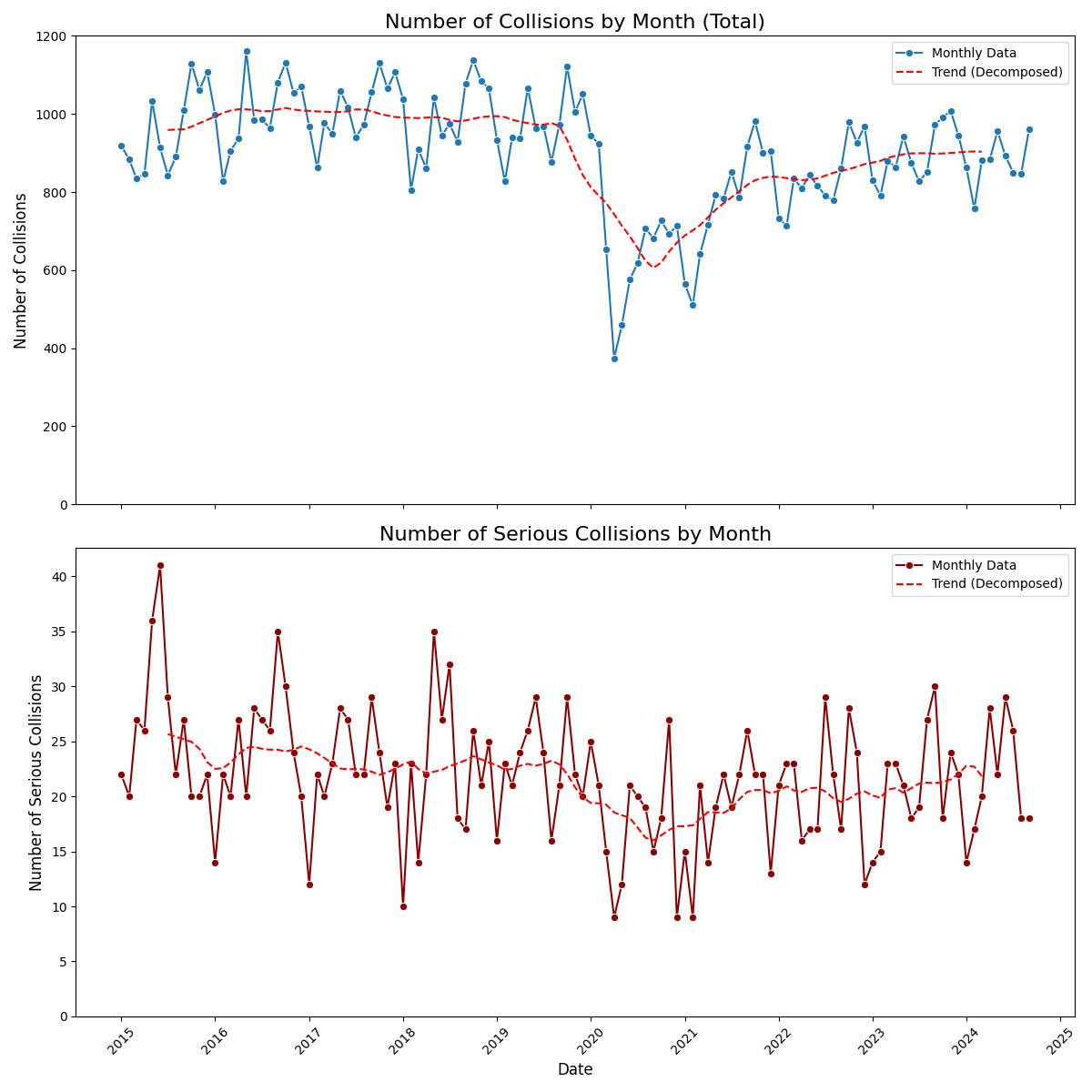}
\caption{Time-Series: Number of collisions from 2015 to 2024 (total / severe only)}
\label{fig:timeseries}
\end{figure}

By overlaying monthly collision data for each year in Figure 3, a distinct cyclic trend emerges. The highest number of collisions occur in May and October, while the fewest are observed in February. We hypothesize that the elevated collision rates in May and October are driven by increased tourism during these months which leads to higher traffic volumes and, consequently, a greater likelihood of accidents. This consistent pattern underscores the significant seasonal variation in traffic accidents. Repost on tourism in Montgomery can be found here~\citep{ref-tourism}.

\begin{figure}
\centering
\includegraphics[width=\textwidth]{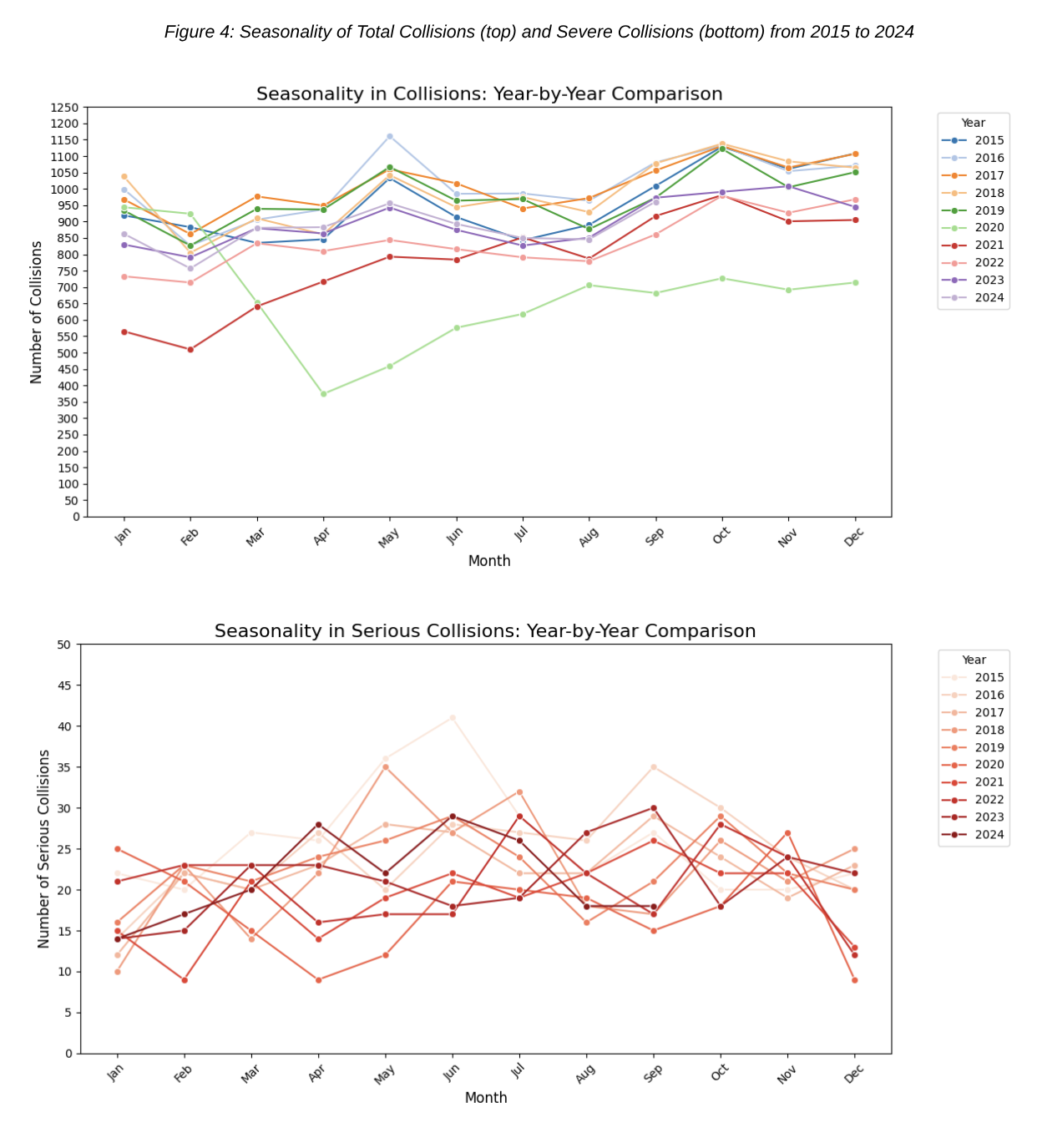}
\caption{Seasonality of Total Collisions (top) and Severe Collisions (bottom) from 2015 to 2024}
\label{fig:seasonality}
\end{figure}

\begin{figure}[H]
\centering
\includegraphics[width=\textwidth]{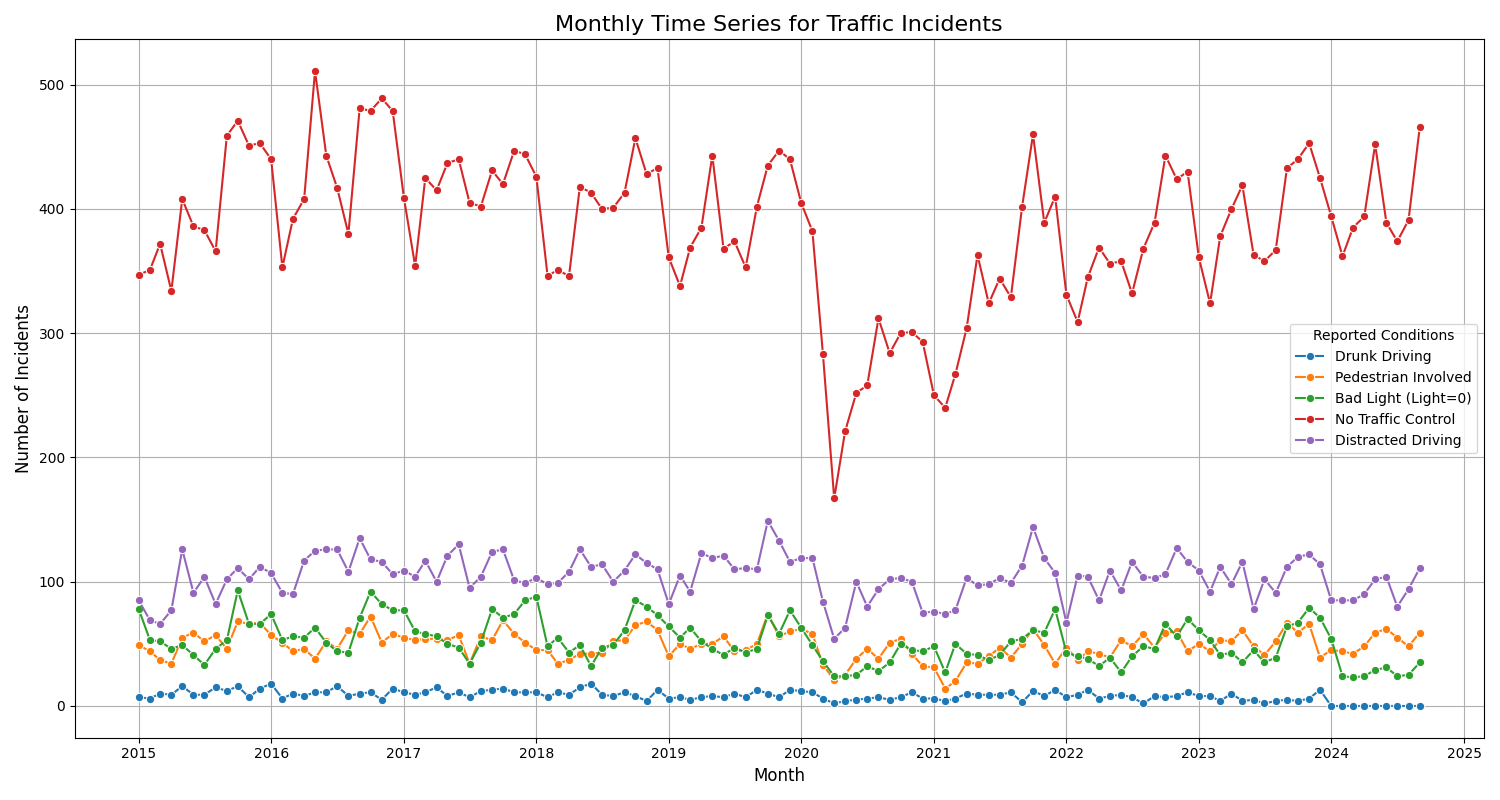}
\end{figure}

Montgomery County, approximately 106,000 traffic crashes were reported, with 2.17\% resulting in fatalities or serious injuries. Of these, 43\% occurred at locations without traffic control. Driver distraction accounted for 11\% of crashes, poor lighting for 5.7\%, 5.5\% involved pedestrians, and 0.9\% involved drivers under the influence of alcohol. Figure 5 illustrates the temporal variation in collision frequencies under different conditions, showing that the seasonal frequency of collisions under poor lighting conditions is smaller compared to those involving driver distraction.

\subsection{Dense areas for different types of crashes}

It is reasonable to assume spatial patterns in traffic collisions. To generate a heatmap of collisions, we applied the non-parametric Spatial Kernel Density Estimation (KDE) technique to approximate the density of collisions under various circumstances. As expected, the estimated density of both total and severe crashes is concentrated in densely populated areas of the county, including Wheaton--Glenmont (in the southern part of the county, near Washington, DC) and the Gaithersburg--Germantown area (a densely populated northern region). Another pattern (fig. 6) is that the most dense areas for collisions with pedestrians are located in most populated towns of the county. Figure 7 also reveals interesting spatial patterns on both plots: most collisions with animals occurred in the northern part (near Germantown) as this area is surrounded by forests and national parks.

\begin{figure}[H]
\centering
\includegraphics[width=\textwidth]{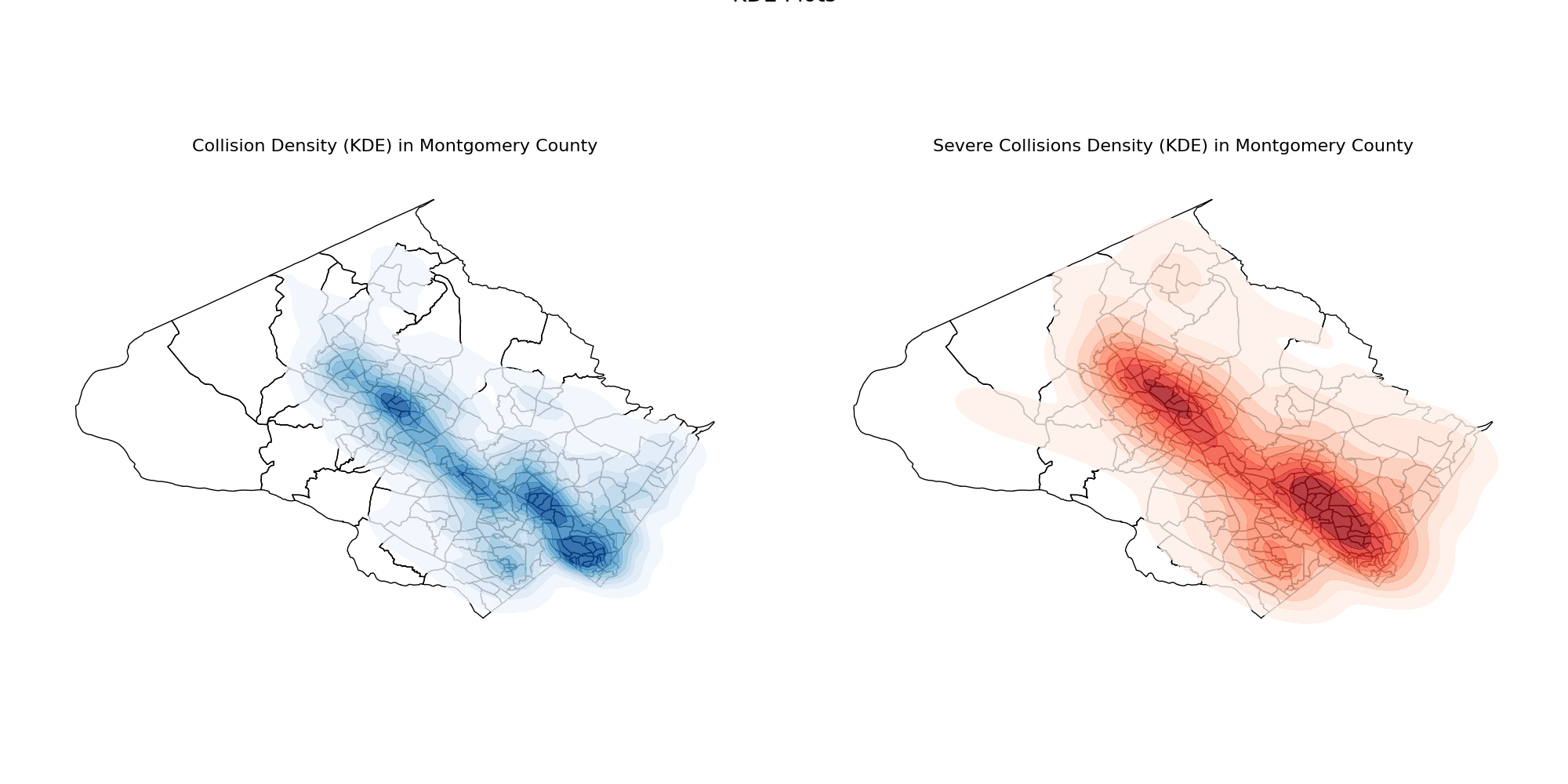}
\caption{Spatial KDE map for collisions (total/severe)}
\label{fig:kde_total_severe}
\end{figure}

\begin{figure}[H]
\centering
\includegraphics[width=\textwidth]{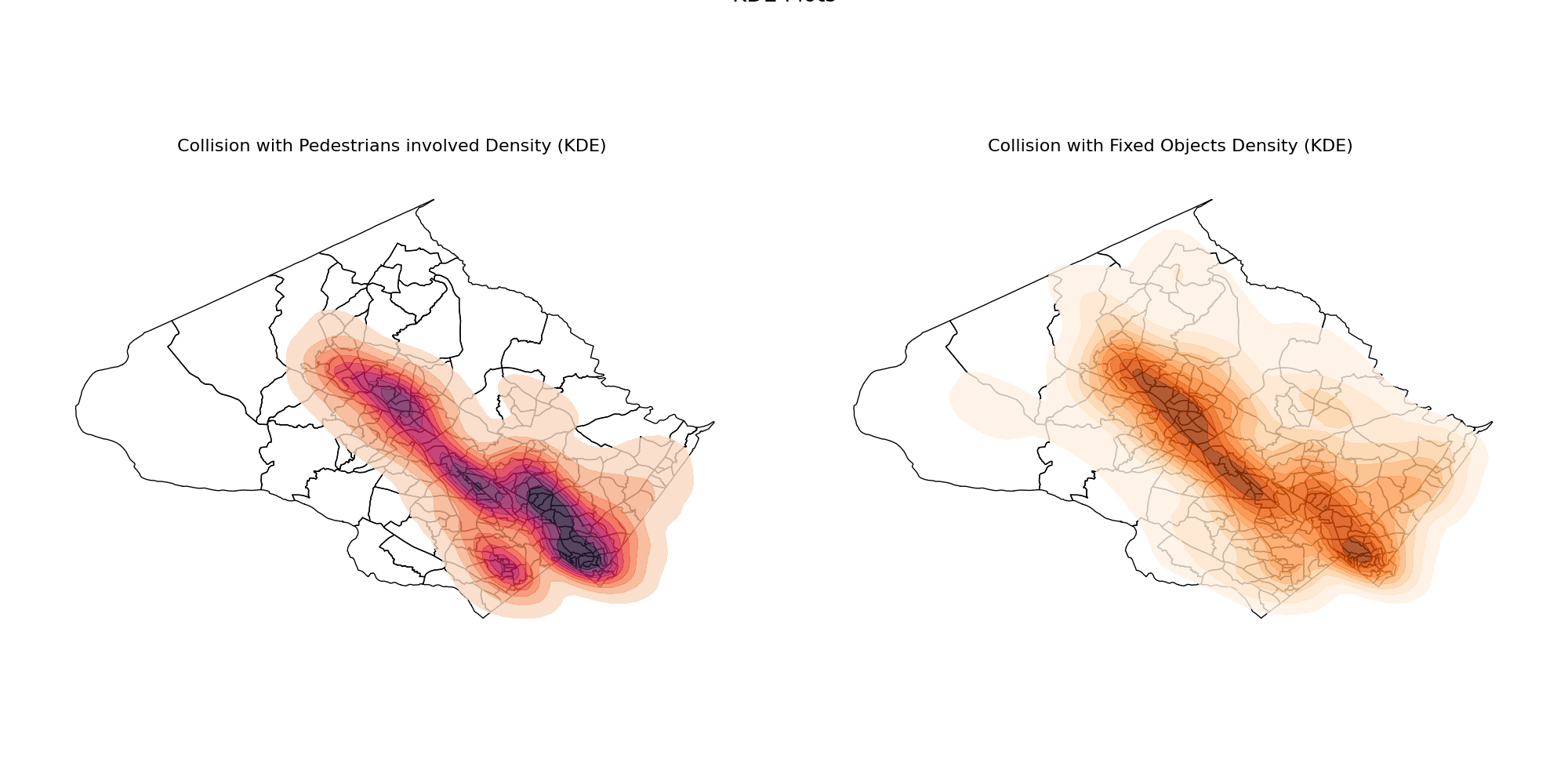}
\caption{Spatial KDE map for collisions with pedestrians and with fixed objects}
\label{fig:kde_pedestrians_fixed}
\end{figure}

\begin{figure}[H]
\centering
\includegraphics[width=\textwidth]{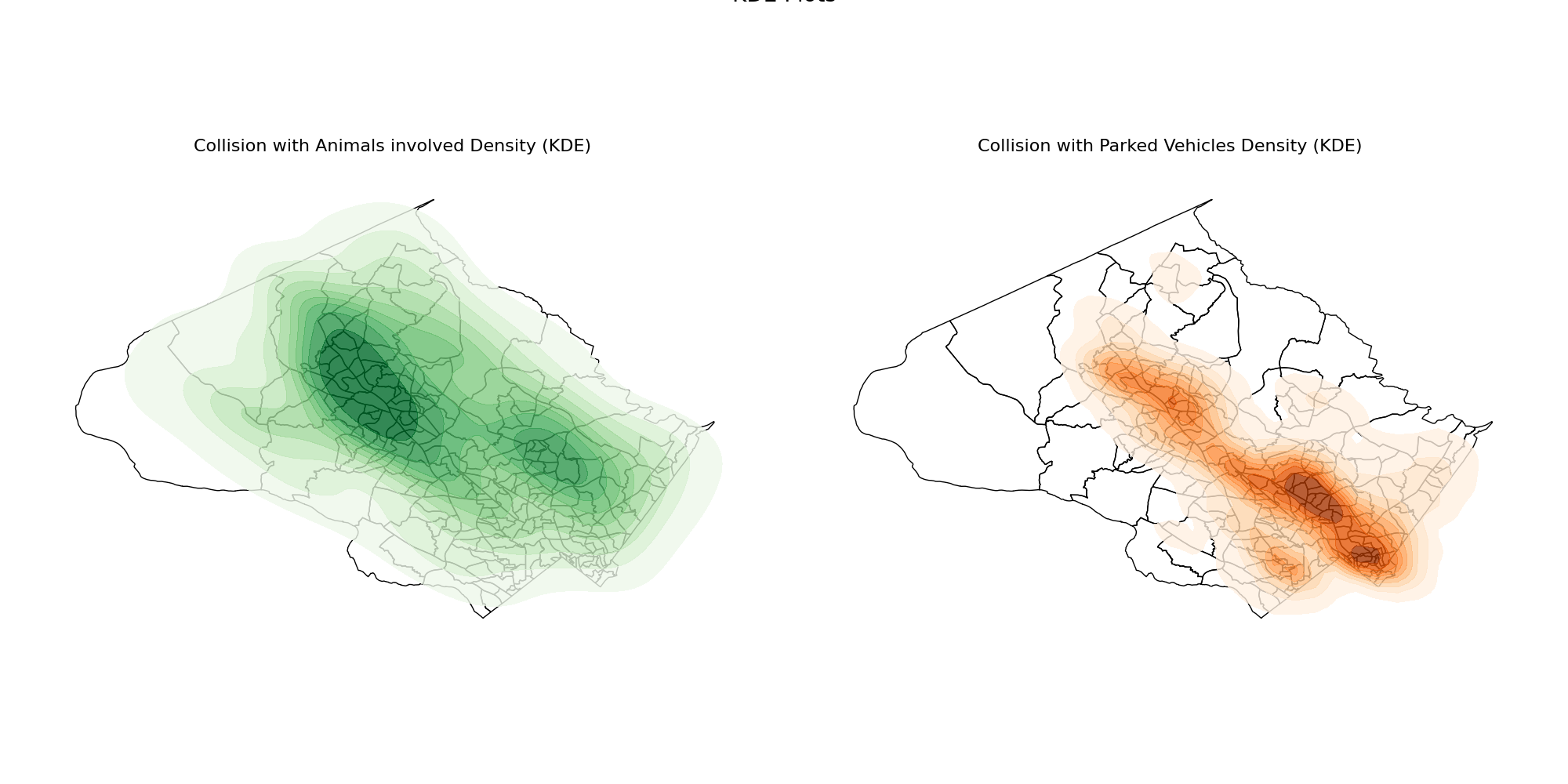}
\caption{Spatial KDE map for collisions with animals and with parked vehicles}
\label{fig:kde_animals_parked}
\end{figure}

\begin{figure}[H]
\centering
\includegraphics[width=\textwidth]{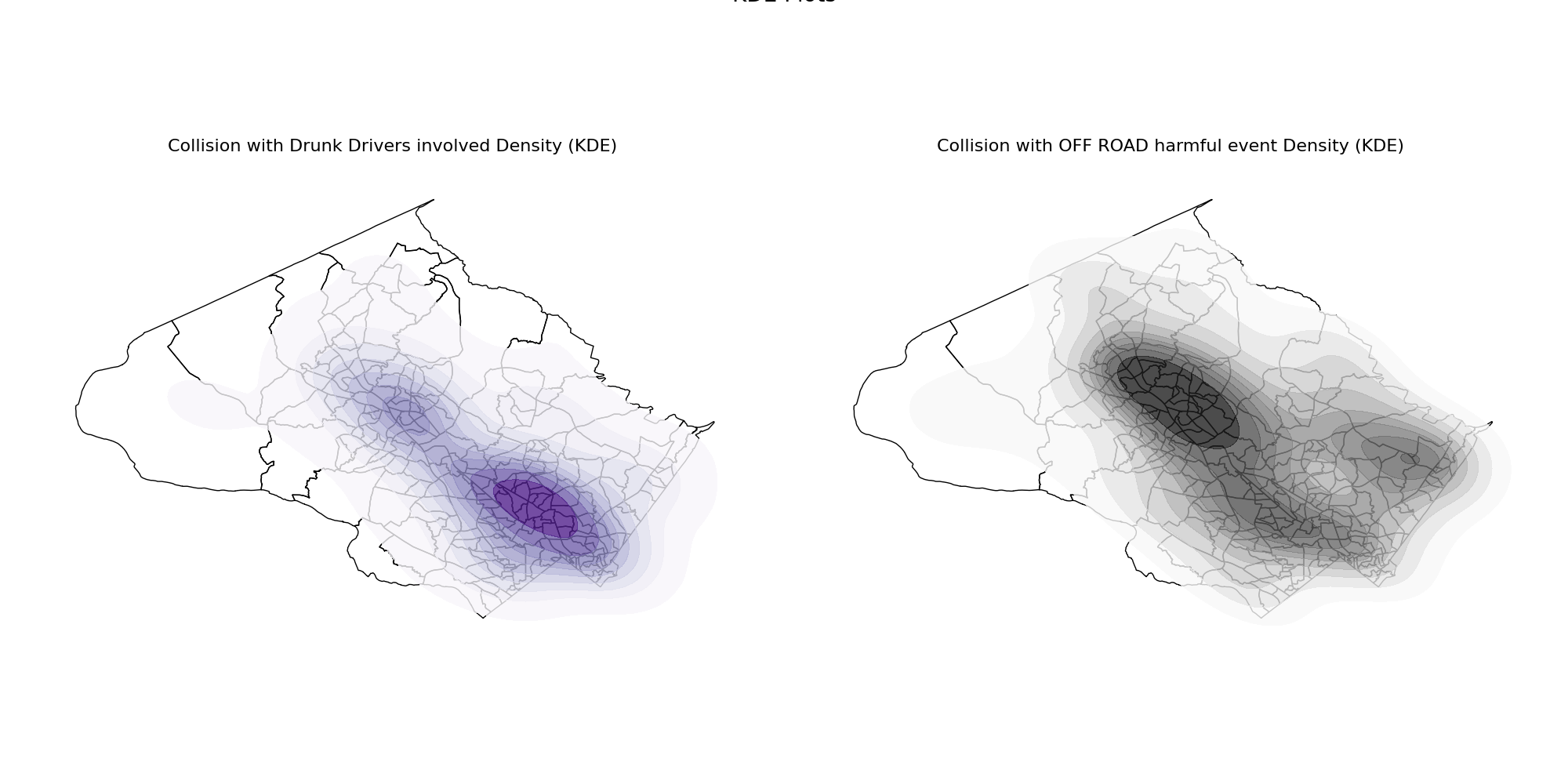}
\caption{Spatial KDE map for collisions with pedestrians and with fixed objects}
\label{fig:kde_pedestrians_fixed2}
\end{figure}

\subsection{Spatial Analysis with Local Moran and LISA}

\subsubsection{LISA for counts of collision within census tracts}

To analyze spatial autocorrelation in traffic collisions, each collision location was mapped to its corresponding census tract within Montgomery County. All subsequent analyses were conducted at the census tract level. We first computed the global Moran's I statistic to test the null hypothesis of spatial randomness. Utilizing a spatial weights matrix constructed with the k-nearest neighbors (KNN) method (k = 10) and a significance level of 0.05, the Moran's I value was 0.11 (p-value = 2.4e-5). This result rejected the null hypothesis, indicating significant positive spatial autocorrelation, where census tracts with high collision counts tend to cluster spatially.

\begin{figure}[H]
\centering
\includegraphics[width=0.6\textwidth]{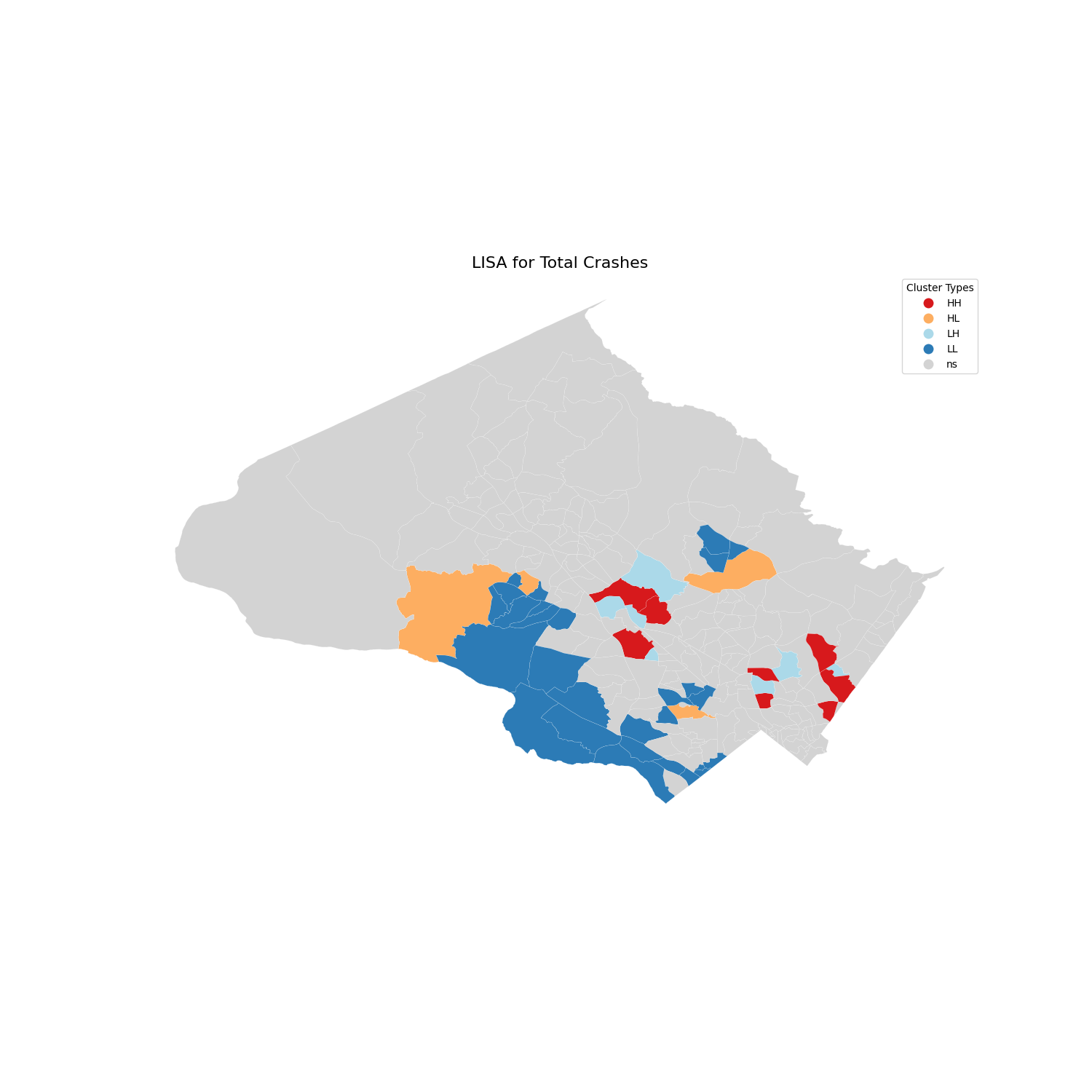}
\caption{LISA clusters of collisions counts (significance level = 0.05)}
\label{fig:lisa_counts}
\end{figure}

\begin{figure}[H]
\centering
\includegraphics[width=0.6\textwidth]{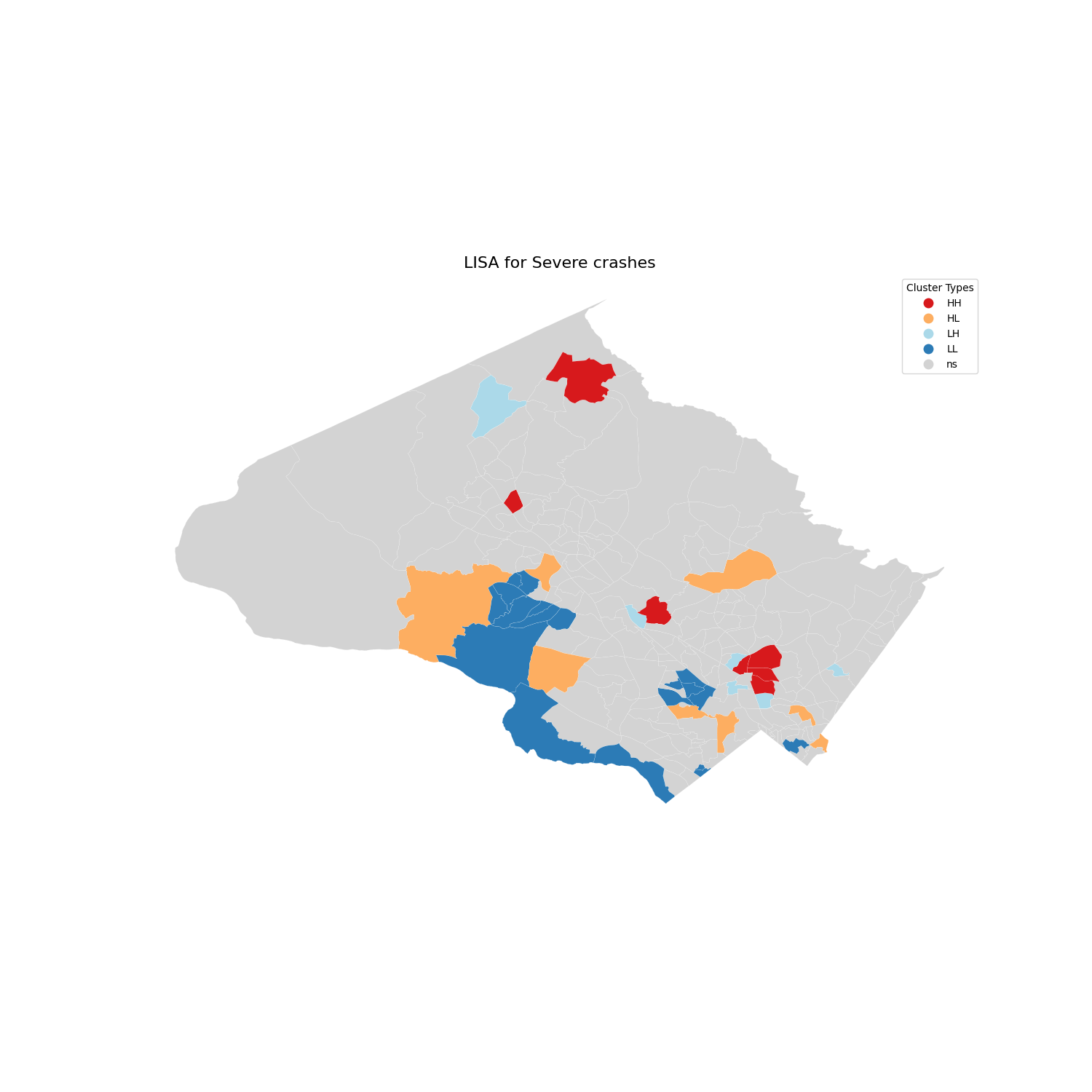}
\caption{LISA clusters of severe collisions counts (significance level = 0.05)}
\label{fig:lisa_severe}
\end{figure}

Once spatial randomness was rejected, the Local Indicators of Spatial Association (LISA) method was employed to examine local spatial patterns of traffic collisions in more detail. LISA provided deeper insights into collision clustering by calculating the Local Moran's I for each census tract, assessing both the significance and the relative value of the target variable. Significant locations were categorized into cluster types: high-high (HH), high-low (HL), low-high (LH), and low-low (LL).

The LISA results for all crashes (Figure 9) and for only severe crashes (Figure 10) indicate that significant cluster locations are similar yet distinct. High-high (HH) clusters for all collisions (Figure 9) are primarily located in densely populated areas of the county, such as Rockville, Wheaton, and Glenmont. In contrast, significant HH clusters for severe collisions (Figure 10) are distributed differently, including areas like Damascus, Germantown, Rockville, Wheaton, and Glenmont. This distinction suggests that while densely populated regions are prone to a higher number of collisions overall, severe collisions also occur in additional areas (Damascus and Germantown) and require further analysis.

\subsubsection{LISA for severity rate among tracts}

An important aspect of collision spatial analysis is assessing the extent of damage inflicted on the involved parties in different areas. In our study, we define this as the "average severity," calculated as the rate of severe collisions (the number of severe collisions divided by the total number of collisions in an area). This metric effectively identifies both high-risk areas with elevated severity rates and relatively safer regions with lower severity rates. However, using rates introduces a challenge: the variance of a proportion depends on its expected value. Consequently, different rates exhibit varying variances, violating the homoscedasticity (constant variance) assumption.

To address this issue, we applied the Empirical Bayes Index (EBI)~\citep{ref-ebi} to stabilize the variance of the severity rates. The EBI method adjusts the estimated rates by incorporating the uncertainty associated with each rate, effectively shrinking them toward the global average proportion in proportion to their variance. Additionally, to handle zero rates and prevent division by zero errors, Laplacian smoothing~\citep{ref-laplace} was employed.

The severity rate for each census tract $i$ is calculated as:
\begin{equation}
\text{severity\_rate}_i = \frac{\text{severe\_crashes}_i + 1}{\text{total\_crashes}_i + 2}
\end{equation}

The standard deviation of the severity rate is:
\begin{equation}
\text{severity\_rate\_std}_i = \sqrt{\frac{(1 - \text{severity\_rate}_i) \times \text{severity\_rate}_i}{\text{total\_crashes}_i + 2}}
\end{equation}

The global severity rate across all $N$ census tracts is:
\begin{equation}
\text{global\_severity\_rate} = \frac{\sum_{i=1}^{N} \text{severe\_crashes}_i + 1}{\sum_{i=1}^{N} \text{total\_crashes}_i + 2}
\end{equation}

The Empirical Bayes Index (EBI) severity for each tract is:
\begin{equation}
\text{EBI\_severity}_i = \frac{\text{severity\_rate}_i - \text{global\_severity\_rate}}{\text{severity\_rate\_std}_i}
\end{equation}

Finally, the standardized EBI severity is:
\begin{equation}
\text{EBI\_severity\_standardized}_i = \frac{\text{EBI\_severity}_i - \mu_{\text{EBI}}}{\sigma_{\text{EBI}}}
\end{equation}

where $\mu_{\text{EBI}}$ and $\sigma_{\text{EBI}}$ are the mean and standard deviation of the EBI severity values across all census tracts.

To address this issue, we applied the Empirical Bayes Index (EBI)~\citep{ref-ebi} to stabilize the variance of the severity rates. The EBI method adjusts the estimated rates by incorporating the uncertainty associated with each rate, effectively shrinking them toward the global average proportion in proportion to their variance. (figure 12) Additionally, to handle zero rates and prevent division by zero errors, Laplacian smoothing~\citep{ref-laplace} was employed.

\begin{figure}[H]
\centering
\includegraphics[width=0.7\textwidth]{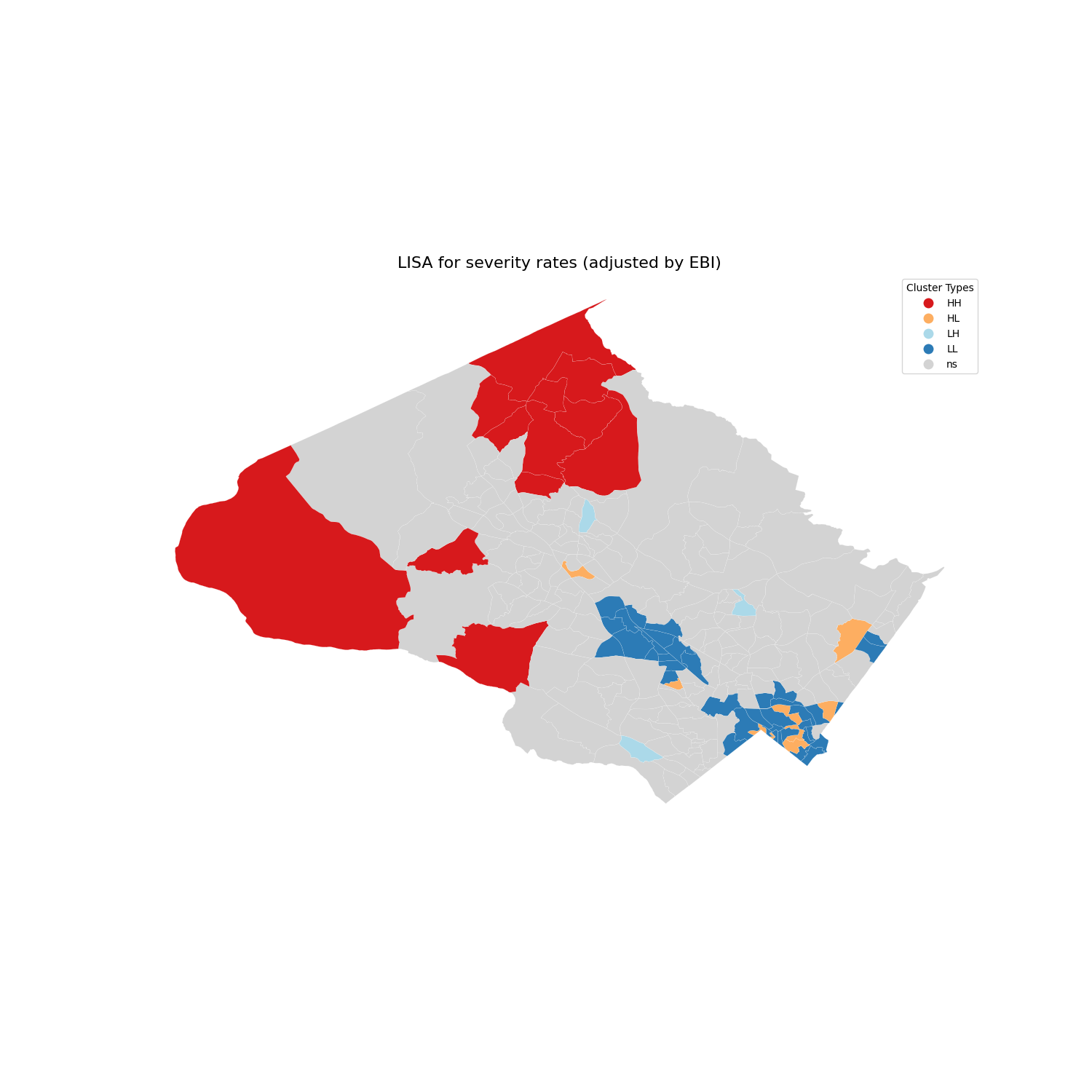}
\caption{LISA for EBI of severity rate. (significance level = 0.05)}
\label{fig:lisa_ebi}
\end{figure}

The LISA clusters for the adjusted severity rate, illustrated in Figure 12, reveal that hotspots (HH clusters) are not concentrated around the county's most populated towns. Instead, these hotspots are primarily located in less populated rural areas such as Damascus and Poolesville. In contrast, the most populated areas are predominantly classified within Low-Low clusters, exhibiting lower severity rates. These hotspots signify regions with a higher risk of severe road crashes relative to the county's overall severity rate, while the Low-Low clusters indicate safer, more densely populated regions. This spatial distribution suggests two plausible hypotheses for the observed pattern:

\begin{itemize}
\item \textbf{Infrastructure and Traffic Control Deficiencies:} These rural areas may be more hazardous due to poorer road conditions, inadequate lighting, insufficient traffic control measures, and more aggressive driving practices.
\item \textbf{Underreporting of Minor Incidents:} The remoteness of these areas from police stations may result in the underreporting of minor collisions, causing only severe crashes to be recorded and thereby increasing the apparent severity rate.
\end{itemize}

\subsubsection{LISA for different types of collisions}

The spatial patterns of traffic collisions vary significantly when analyzed by different types, such as alcohol-related crashes, pedestrian involvement, off-road incidents, and distracted driving. Figure 13 (1 row, 1 col) illustrates that densely populated areas, including Rockville, Wheaton, and Glenmont, form significant high-high (HH) clusters for collisions involving pedestrians. In contrast, rural areas are primarily classified as low-low (LL) clusters, indicating significantly fewer pedestrian-involved collisions.

For alcohol-related crashes (Figure 13, Row 1, Column 2), the LISA analysis identified a single significant HH cluster spanning the Wheaton and Glenmont areas. For collisions involving parked vehicles (Figure 13, Row 3, Column 1), the results are as expected, with significant clusters concentrated in the county's most populated towns. In the case of collisions caused by distracted drivers, significant clusters are observed in tracts spanning Rockville, Wheaton, Glenmont, and also Poolesville, which notably do not appear as significant clusters in other LISA maps.

\begin{figure}
\centering
\begin{tabularx}{\textwidth}{CC}
\includegraphics[width=0.45\textwidth]{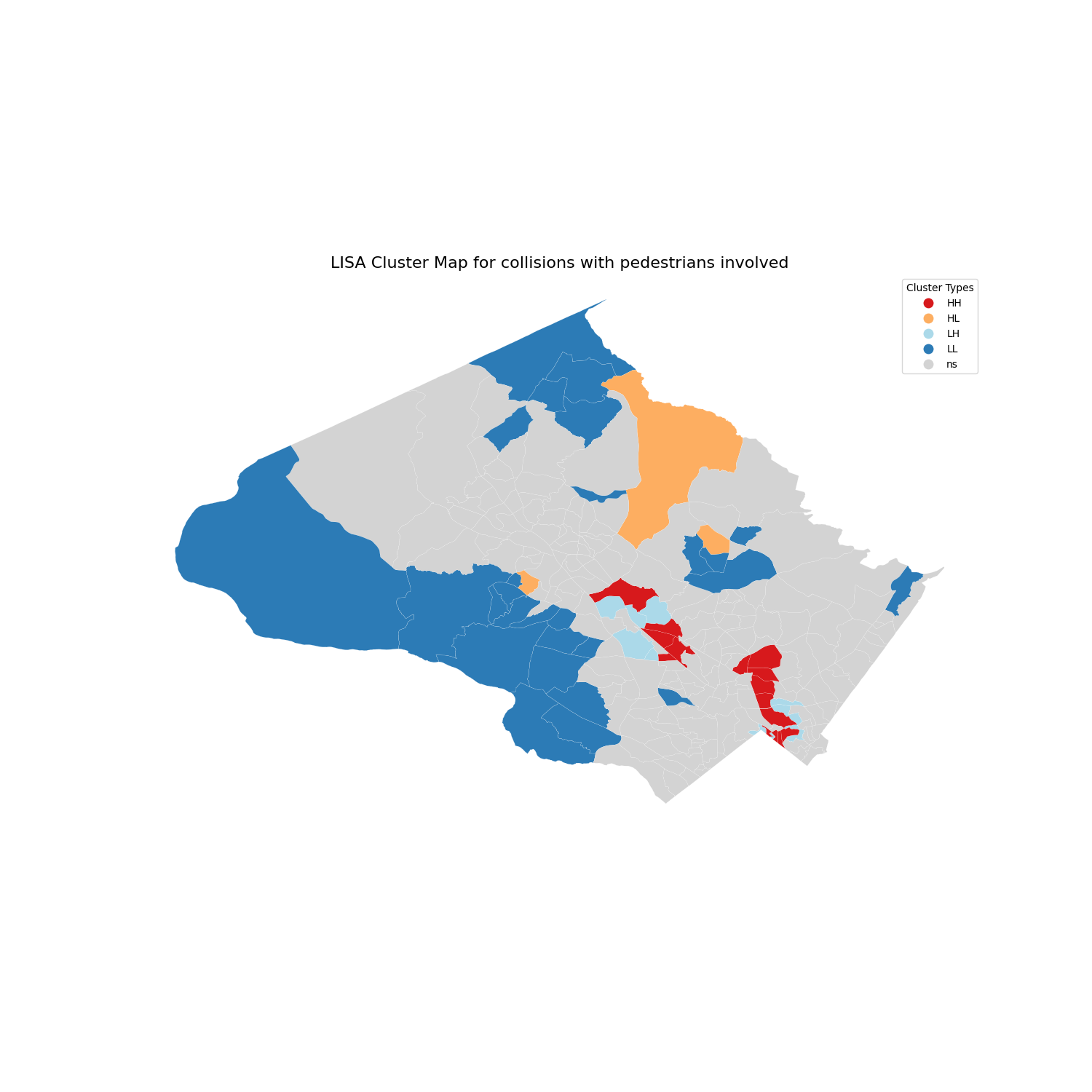} & 
\includegraphics[width=0.45\textwidth]{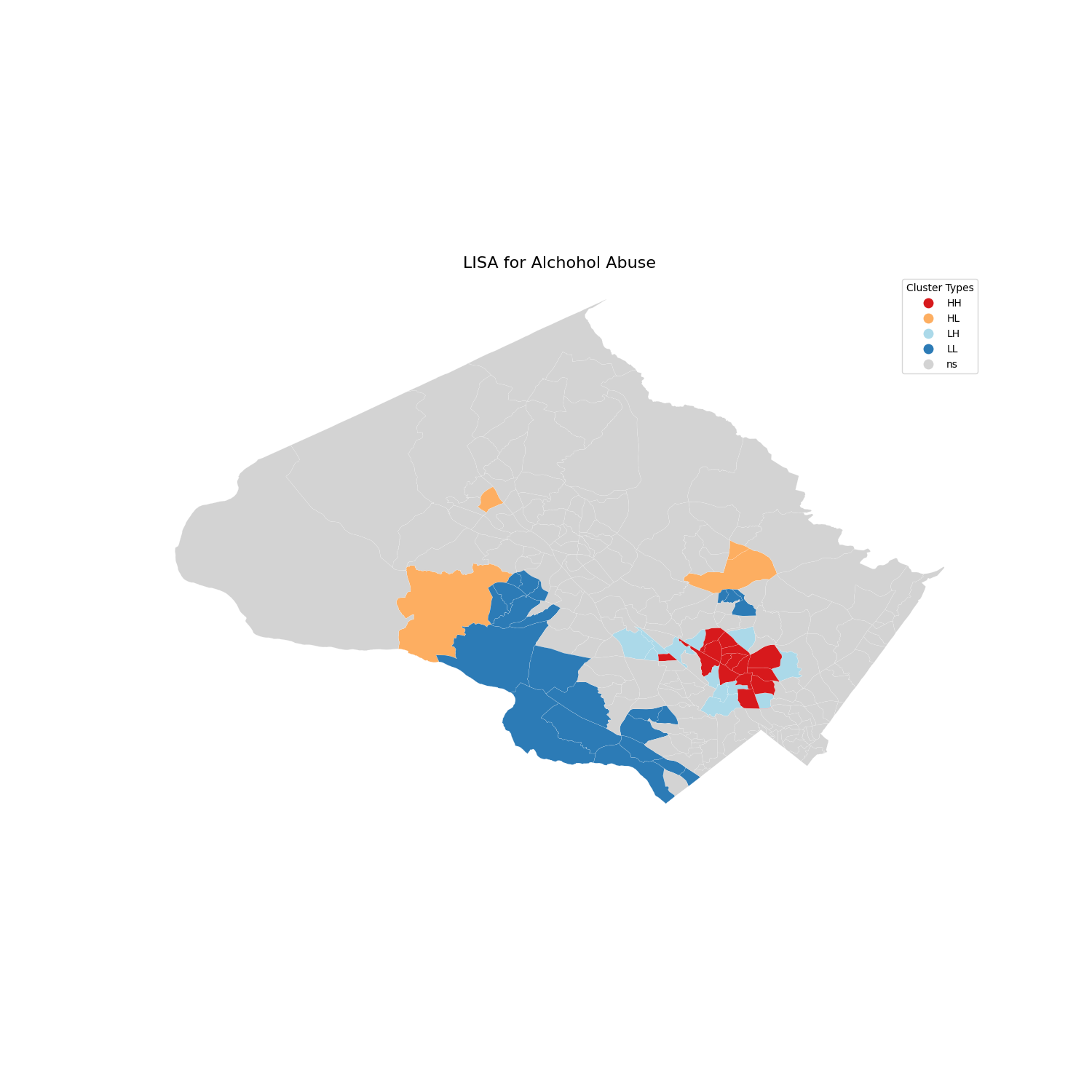} \\
\includegraphics[width=0.45\textwidth]{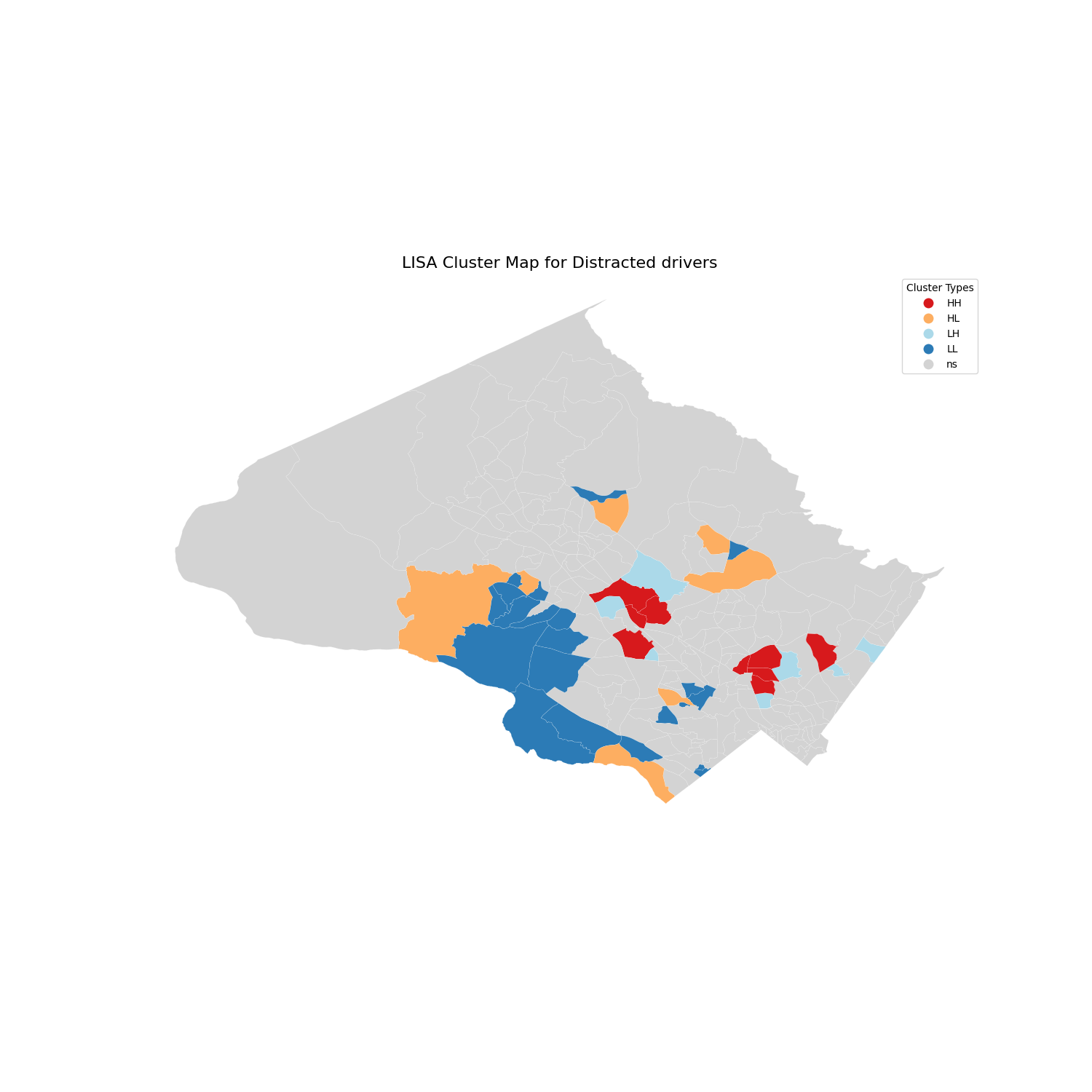} & 
\includegraphics[width=0.45\textwidth]{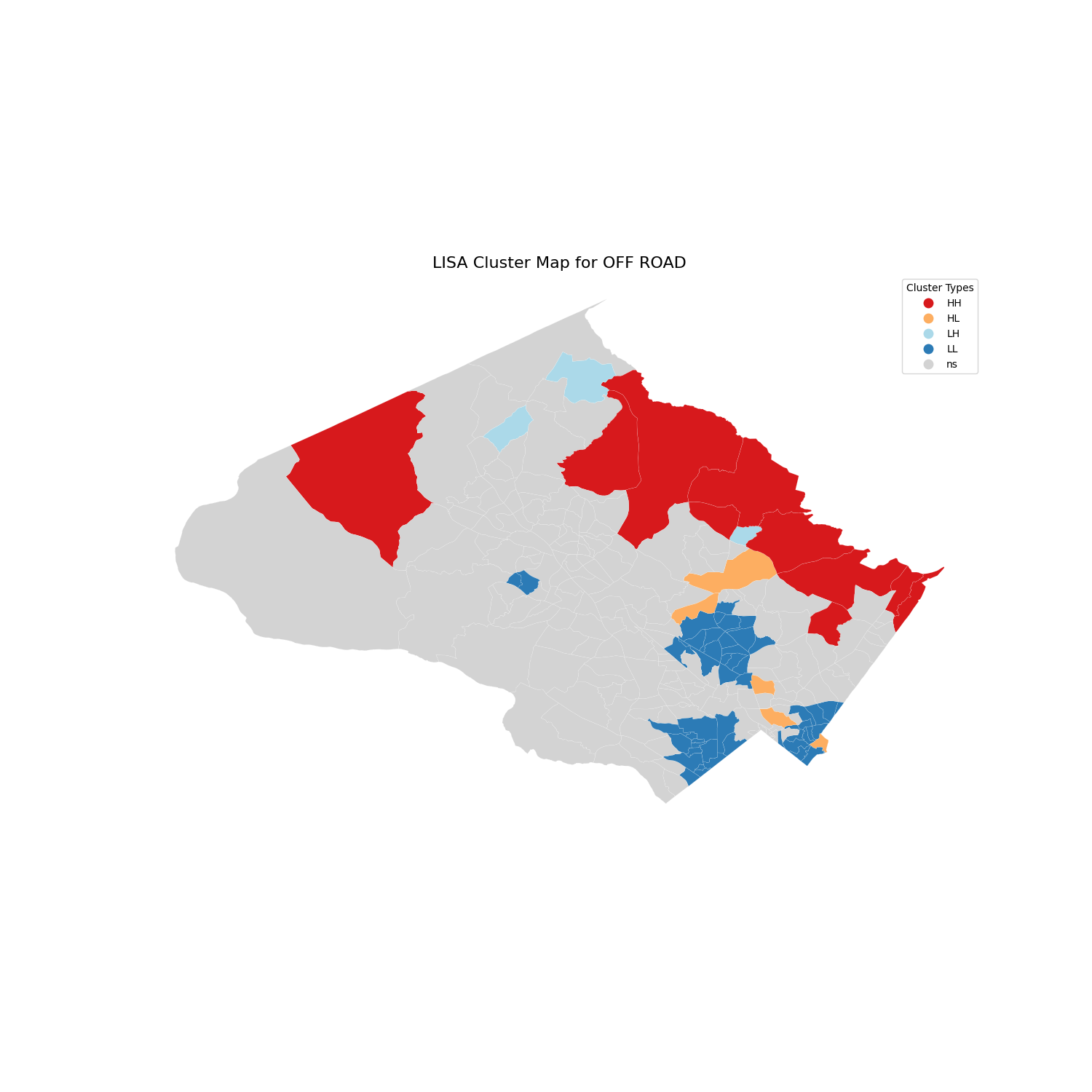} \\
\includegraphics[width=0.45\textwidth]{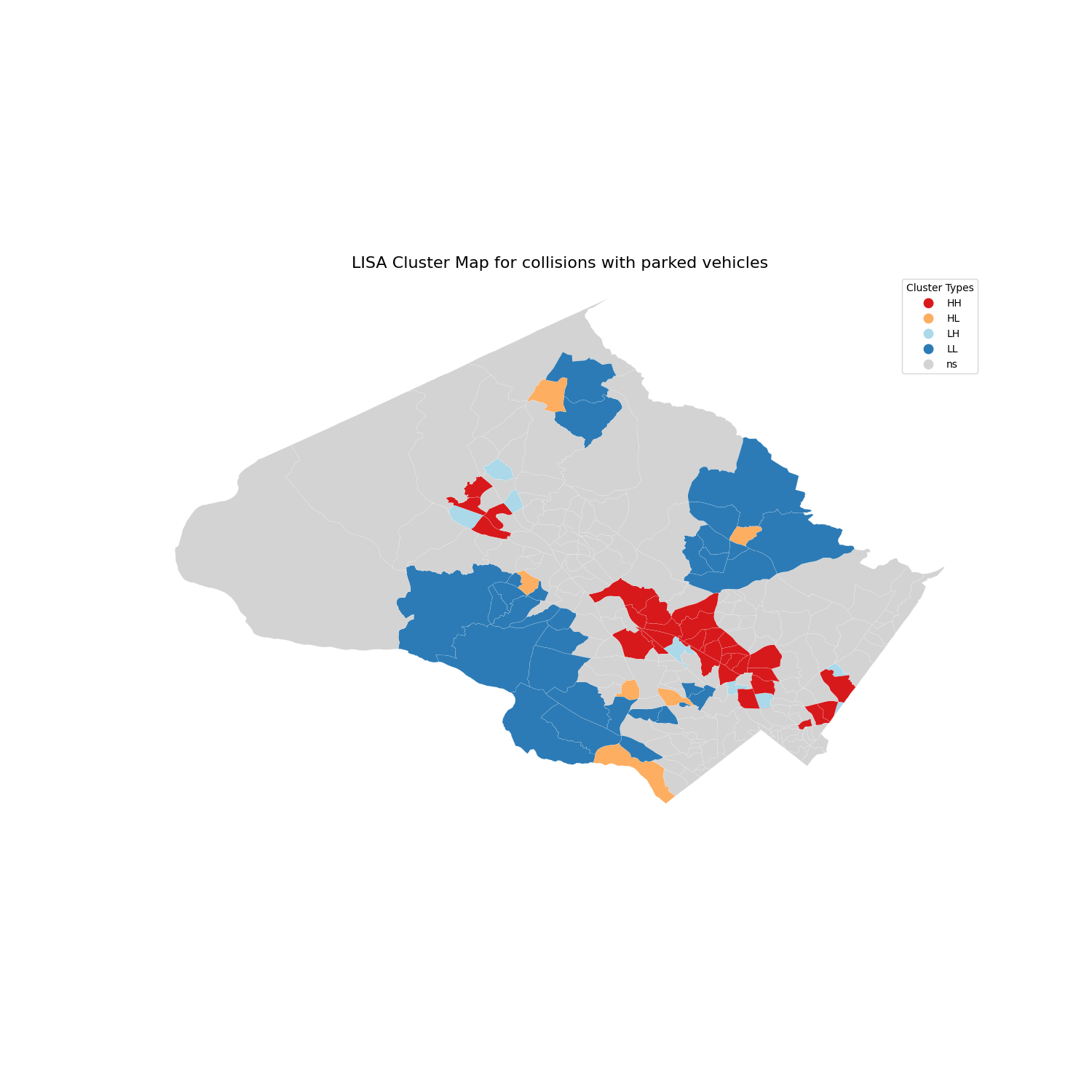} & 
\includegraphics[width=0.45\textwidth]{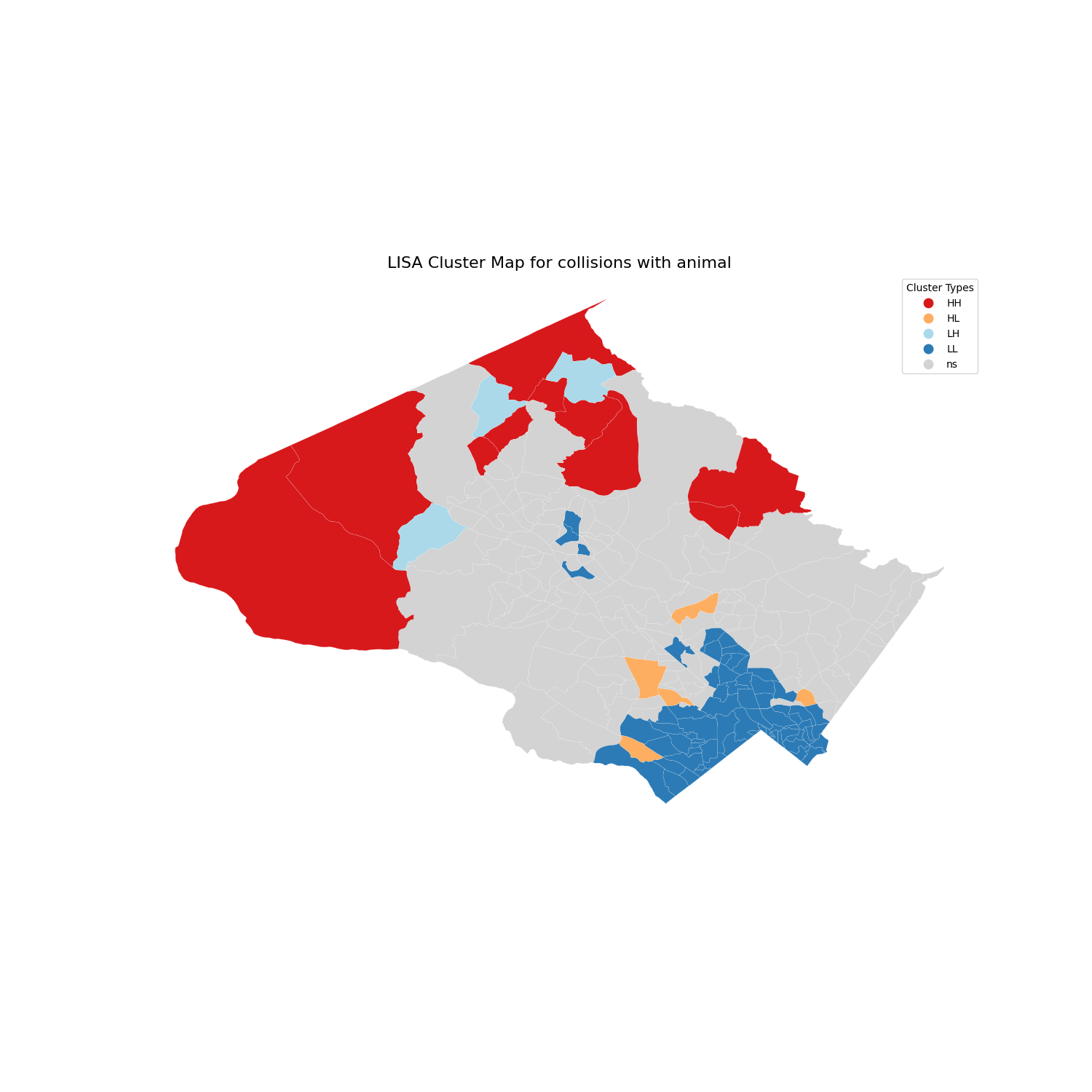}
\end{tabularx}
\caption{LISA for crashes with specific factors or circumstance}
\label{fig:lisa_types}
\end{figure}

Crashes involving animals (Figure 13, Row 2, Column 2) show a distinct pattern compared to total collisions, with significant HH clusters located in rural, less populated areas near forests and national parks. A similar spatial pattern is observed for off-road collisions (Figure 13, Row 3, Column 2), which are also concentrated in the county's sparse, rural regions.

\subsubsection{Crashes points distribution for significantly dense areas}

To identify critical areas for traffic safety analysis, we focused on census tracts within Montgomery County's most populated towns that exhibited both a relatively high number of collisions and significant spatial autocorrelation. As illustrated in Figure 14, crash points within these selected tracts are marked in gray, whereas severe crashes are denoted by red markers. Further examination of these densely collision-prone locations, shown in Figure 15, reveals that most collisions occur at intersections.

\begin{figure}[H]
\centering
\begin{tabularx}{\textwidth}{CC}
\includegraphics[width=0.48\textwidth]{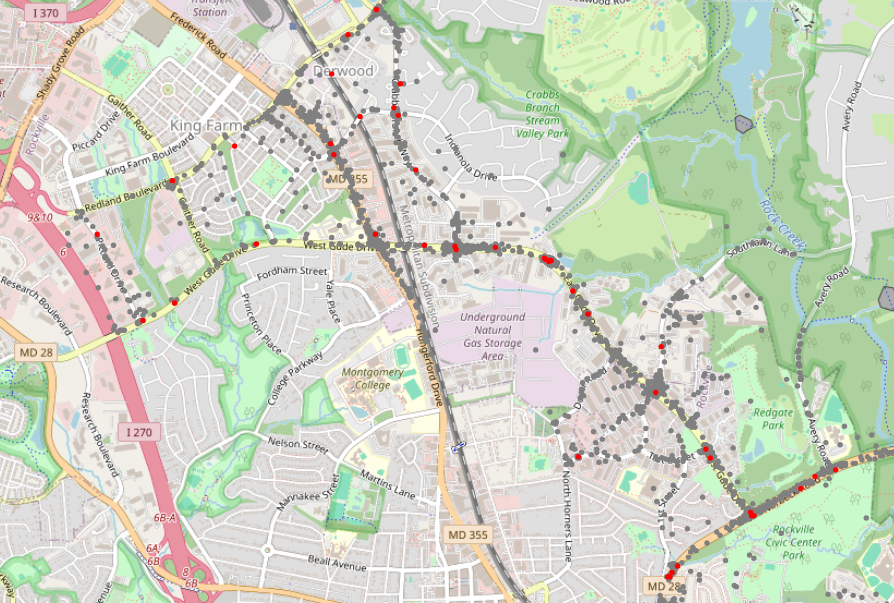} & 
\includegraphics[width=0.48\textwidth]{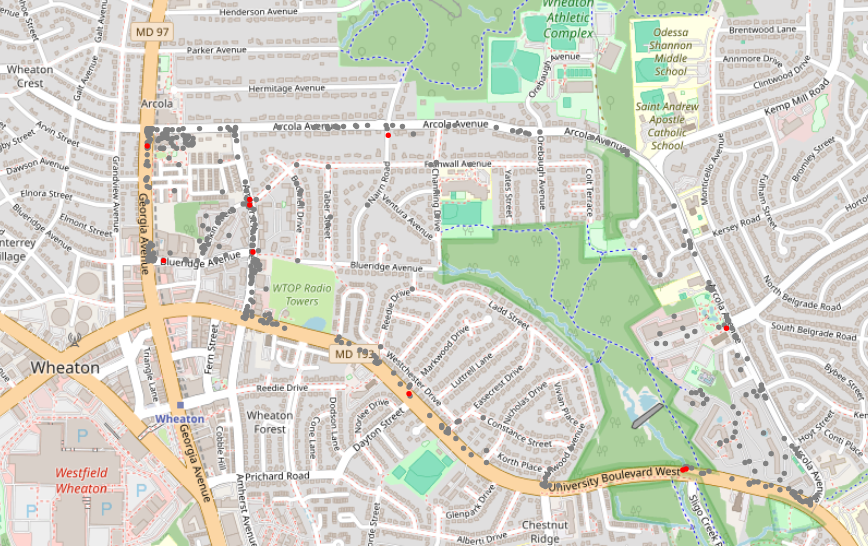} \\
\includegraphics[width=0.48\textwidth]{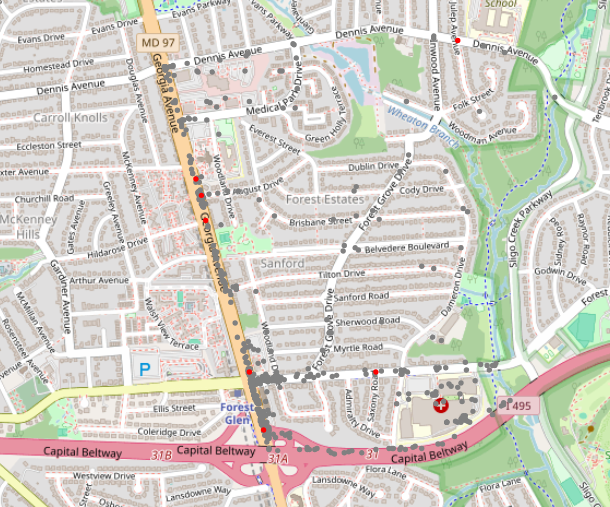} & 
\includegraphics[width=0.48\textwidth]{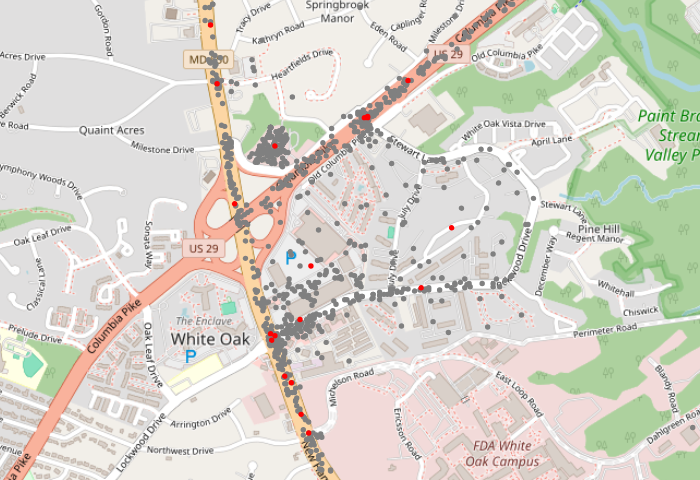}
\end{tabularx}
\caption{Distribution of crash points in significant areas (first row - Rockville area and Wheaton, second row - Montgomery hills and White Oak)}
\label{fig:crash_distribution}
\end{figure}

\begin{figure}[H]
\centering
\includegraphics[width=0.7\textwidth]{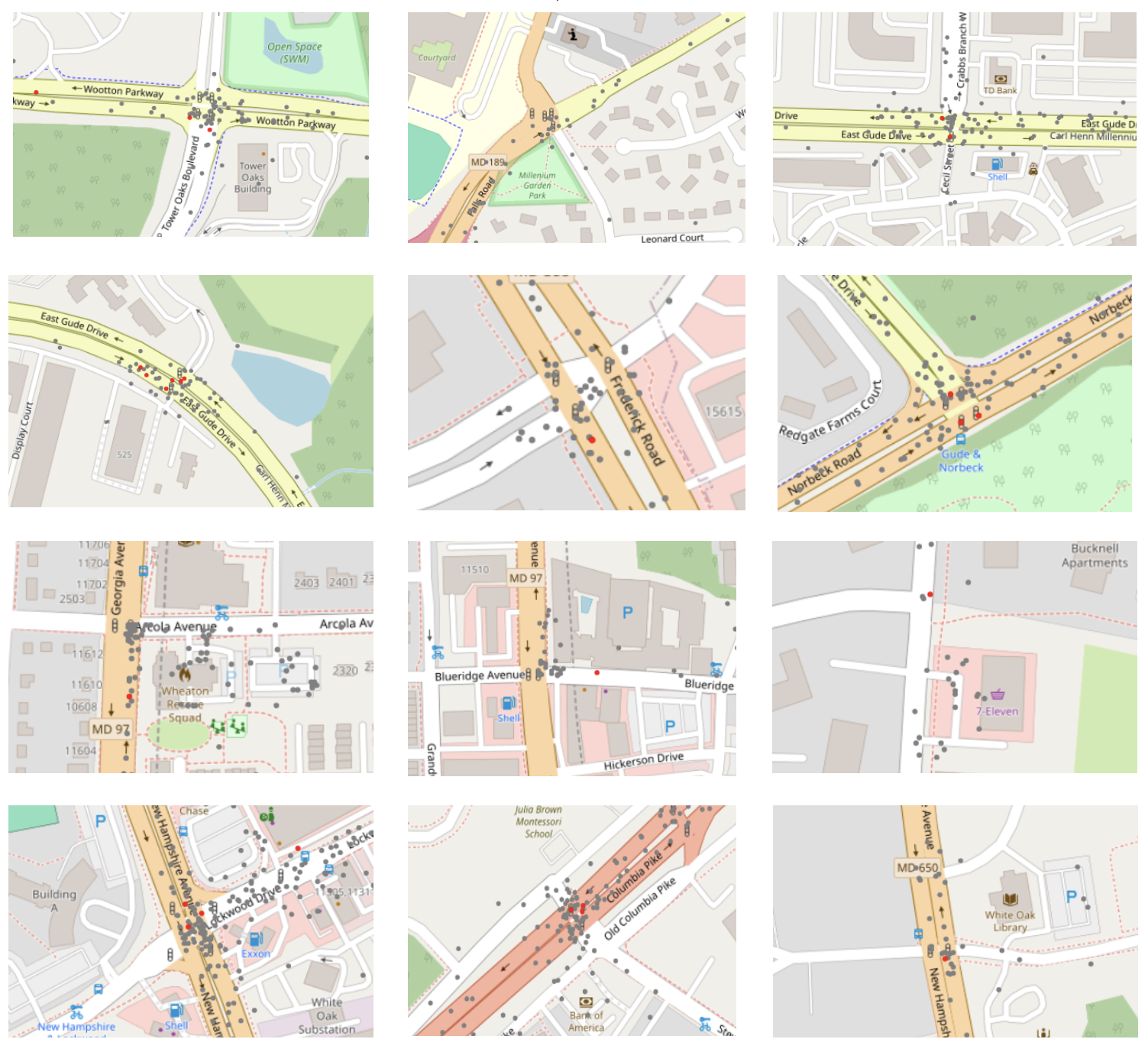}
\caption{Distribution of crash points in significant areas (first row - Rockville area and Wheaton, second row - Montgomery hills and White Oak)}
\label{fig:crash_details}
\end{figure}

\section{Discussion}

This study found that there is enough evidence to reject hypotheses about spatial randomness in collisions and that there is a significant positive autocorrelation which means areas with large collision density tend to cluster together close to each other. By looking at a more granular level of individual census tracts and applying LISA, we observed an expected pattern that collisions tend to occur and cluster at most populated areas such as big towns and cities or loaded interstate highways. Similar pattern of clustering near populated towns is observed when considering only severe collisions. Tracts with significant autocorrelation and relatively high number of collisions: 70390, 700903, 70121, 70150, 701001

In contrast, when we considered not the counts of collisions, which depend heavily on the population of a tract, but the adjusted severity rate, LISA uncovered significant positively autocorrelated tracts located in more rural and less populated areas such as Poolesville and Damascus as tracts with relatively high severity rates for collisions. This observation suggests two plausible explanations for the observed pattern: these rural areas may be more hazardous due to poorer road conditions, inadequate lighting, insufficient traffic control measures, and more aggressive driving practices, or the remoteness of these areas from police stations (see Figure 17) may lead to underreporting of minor collisions, resulting in only severe crashes being recorded and thereby inflating the apparent severity rate. Tracts with significant autocorrelation and relatively high severity rate: 700500, 700204, 700207, 700209

\begin{figure}[H]
\centering
\includegraphics[width=0.7\textwidth]{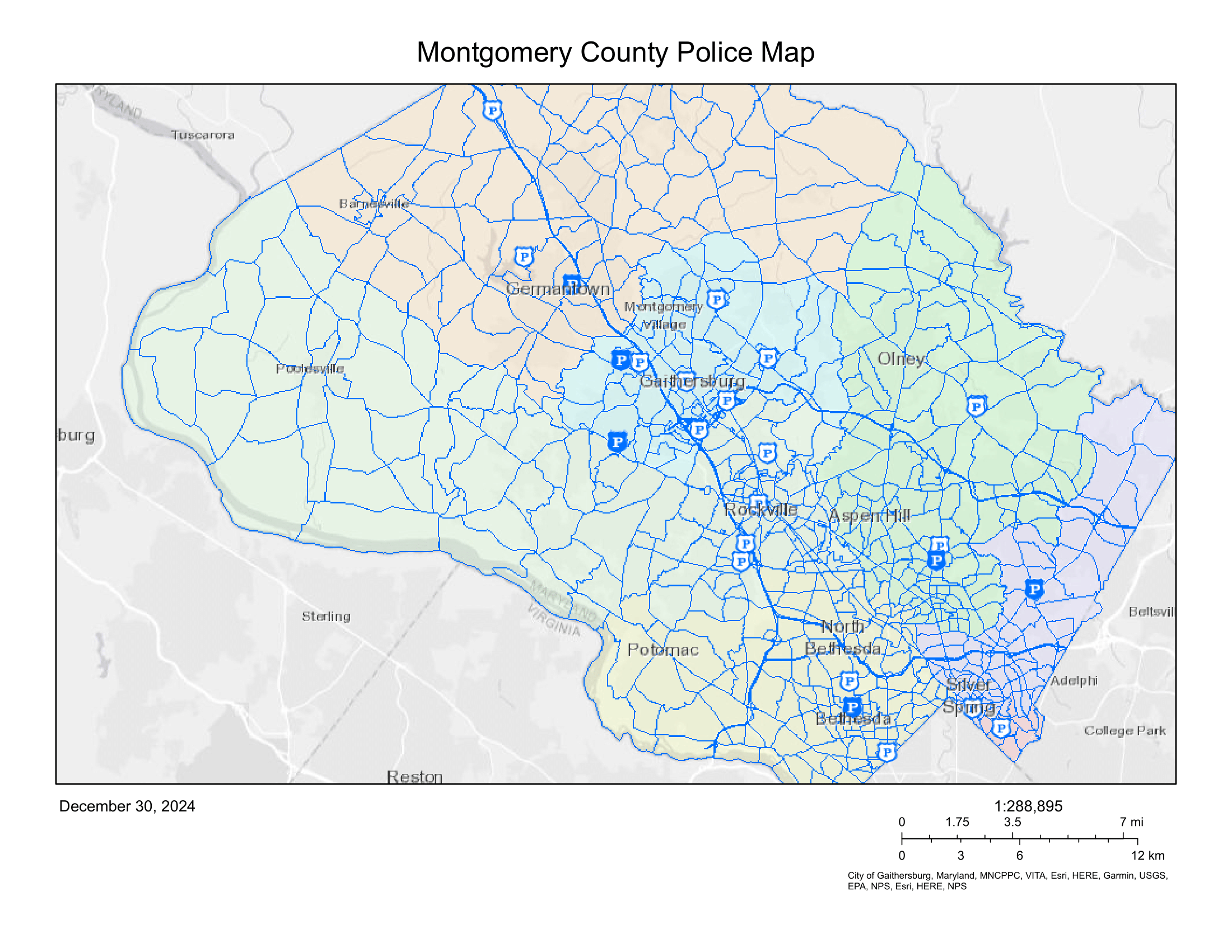}
\caption{Locations of Police Stations in Montgomery, MD}
\label{fig:police_stations}
\end{figure}

Certain crash conditions or contributing factors are modifiable and may be mitigated through targeted interventions. Analyzing crashes by specific conditions (Figure 14, 1st row, 2nd col) reveals a notable pattern for alcohol-related collisions. The analysis identified a single significant high-high (HH) cluster located in the Wheaton-Glenmont area, surrounded by significant low-low (LL) clusters. The factors contributing to the elevated incidence of alcohol-related collisions in this region remain unclear and require further investigation to inform effective intervention strategies.

Collisions involving animals tend to form significant clusters in relatively rural areas (Figure 14, 2st row, 2nd col) adjacent to forests and national parks. Although it may be impractical to eliminate these incidents entirely, certain interventions can mitigate their occurrence. For example, implementing reduced speed limits in zones where animals are most likely to cross the road can diminish collision risk and enhance overall safety or by constructing green bridges to enable wildlife to cross roads without having to negotiate traffic.

Collisions involving vehicles leaving the roadway (off road type) exhibit a similar spatial pattern of significant clusters, primarily concentrated in rural areas located in the eastern region of the county. This distribution likely corresponds to the presence of high-speed roads and highways in these areas.

Investigating areas with significant crash clustering (Figure 16), we observe that in densely populated regions, collisions frequently occur at intersections and road segments with pedestrian crossings, consistent with findings by Dezman et al. in Baltimore.

\section{Limitations}
The study is subject to several limitations that warrant consideration. Firstly, the reliance on police-reported data may introduce biases due to human error, misreporting, or omissions, and some collisions might not have been reported, potentially affecting the accuracy of the findings. Secondly, the analysis does not incorporate traffic flow or road network load data, which could enhance the estimation of accident probabilities across different road segments. Lastly, while the study identifies high-risk areas, it does not provide specific intervention strategies to reduce motor vehicle crash damage and fatalities. These limitations should be considered when interpreting the results and planning future research.

\section{Conclusion}
Identifying dangerous zones prone to motor vehicle crashes is a critical step toward developing a sustainable policy framework for reducing the damage caused by traffic accidents. This study demonstrates how spatial autocorrelation analysis and kernel density estimation can be applied to identify clusters with a significant concentration of crashes and to explore the spatial distribution of various modifiable risk factors, such as alcohol abuse, distracted driving, and collisions with animals. The findings from this research can serve as a guide for targeted interventions by policymakers, traffic planners, and other stakeholders to enhance road safety across the United States.

\section*{Data Availability}
The data supporting the findings of this study are publicly available at \url{https://catalog.data.gov/dataset/crash-reporting-incidents-data}.

\section*{Acknowledgments}
The author acknowledges the Montgomery County Police Department and Maryland State Police for providing the crash data used in this analysis.

\section*{Conflict of Interest}

\end{document}